\documentclass[pra,twocolumn,showpacs,floatfix]{revtex4}
\usepackage{graphicx}
\begin{document}

\title{Mixtures of Bose gases under rotation}
\author{S. Bargi, J. Christensson, G. M. Kavoulakis,
and S. M. Reimann}
\affiliation{Mathematical Physics, Lund Institute of
Technology, P.O. Box 118, SE-22100 Lund, Sweden}
\date{\today}

\begin{abstract}

We examine the rotational properties of a mixture of two Bose
gases. Considering the limit of weak interactions between the
atoms, we investigate the behavior of the system under a fixed
angular momentum. We demonstrate a number of exact results
in this many-body system.

\end{abstract}
 \maketitle

One of the many interesting aspects of the field of cold atoms
is that one may create mixtures of different species. The
equilibrium density distribution of the atoms is an interesting
problem by itself, since the different components may coexist,
or separate, depending on the value of the coupling constants
between the atoms of the same and of the different species. If
this system rotates, the problem becomes even more interesting.
In this case, the state of lowest energy may involve rotation
of either one of the components, or rotation of all the
components. Actually, the first vortex state in cold gases of
atoms was observed experimentally in a two-component system
\cite{Cornell1}, following the theoretical suggestion of
Ref.\,\cite{Holland}. More recently, vortices have also been
created and observed in spinor Bose-Einstein condensates
\cite{Ketterle,Cornell2}. Theoretically, there have been
several studies of this problem \cite{Ho,Ueda1,Ueda2}, mostly
in the case where the number of vortices is relatively large.
Kasamatsu, Tsubota, and Ueda have also given a review of the
work that has been done on this problem \cite{Uedareview}.

In this Letter, we consider a rotating two-component Bose gas
in the limit of weak interactions and slow rotation, where the
number of vortices is of order unity. Surprisingly, a number of
exact analytical results exist for the energy of this system.
The corresponding many-body wavefunction also has a relatively
simple structure.

We assume equal masses $M$ for the two components, and a
harmonic trapping potential $V_t = M (\omega^2 \rho^2 +
\omega_z^2 z^2)/2$, with $\rho^2 = x^2 + y^2$. The trapping
frequency $\omega_z$ along the axis of rotation is assumed to
be much higher than $\omega$. In addition, we consider weak
atom-atom interactions, much smaller than the oscillator energy
$\hbar \omega$, and work within the subspace of states of the
lowest Landau level. The motion of the atoms is thus frozen
along the axis of rotation and our problem becomes
quasi-two-dimensional \cite{Ben}. The relevant eigenstates are
$\Phi_m(\rho,\theta) \varphi_0(z)$, where $\Phi_m(\rho,\theta)$
are the lowest-Landau-level eigenfunctions of the
two-dimensional oscillator with angular momentum $m \hbar$, and
$\varphi_0(z)$ is the lowest harmonic oscillator eigenstate
along the $z$ axis.

The assumption of weak interactions also excludes the
possibility of phase separation in the absence of rotation
\cite{phasesep}, since the atoms of both species reside in the
lowest state $\Phi_{0,0}({\bf r}) = \Phi_0 (\rho, \theta)
\varphi_0(z)$, while the depletion of the condensate due to the
interaction may be treated perturbatively.

We label the two (distinguishable) components of the gas as $A$
and $B$. In what follows the atom-atom interaction is assumed
to be a contact potential of equal scattering lengths for
collisions between the same species and the different ones,
$a_{AA} = a_{BB} = a_{AB} = a$. The interaction energy is
measured in units of $v_0 = U_0 \int |\Phi_{0,0}({\bf r})|^4 \,
d^3r = (2/\pi)^{1/2} \hbar \omega a/a_z$, where $U_0=4 \pi
\hbar^2 a/M$, and $a_0 = (\hbar/M \omega)^{1/2}$, $a_z = (\hbar
/ M \omega_z)^{1/2}$ are the oscillator lengths on the plane of
rotation and perpendicular to it.

If $N_A$ and $N_B$ denote the number of atoms in each
component, we examine the behavior of this system for a fixed
amount of $L$ units of angular momentum, with $0 \le L \le
N_{\rm max}$, where $N_{\rm max} = {\rm max} (N_A, N_B)$. We
use both numerical diagonalization of the many-body Hamiltonian
for small systems, as well as the mean-field approximation.
Remarkably, as we explain in detail below, there is a number of
exact results in this range of angular momenta.

More specifically, when $0 \le L \le N_{\rm min}$, where
$N_{\rm min} = {\rm min} (N_A, N_B)$, using exact
diagonalization of the many-body Hamiltonian, we find that the
interaction energy of the lowest-energy state has a parabolic
dependence on $L$ in this range,
\begin{eqnarray}
  {\cal E}_0 (L)/v_0 = \frac 1 2 N (N-1)
  - \frac 1 2 N L + \frac 1 4 L (L-1),
\label{yrast1}
\end{eqnarray}
with $N=N_A+N_B$. In addition, the lowest-energy state consists
only of the single-particle states of the harmonic oscillator
with $m=0$ and $m=1$. The occupancy of the $m=1$ state of each
component is given by
\begin{eqnarray}
  (N_A)_{m=1} &=& L \frac {N_B - L + 1} {N - 2 L + 2},
\\
  (N_B)_{m=1} &=& L \frac {N_A - L + 1} {N - 2 L + 2},
\label{occupancies}
\end{eqnarray}
while $(N_A)_{m=0} = N_A - (N_A)_{m=1}$, and $(N_B)_{m=0} = N_B
- (N_B)_{m=1}$.

\begin{figure}[t]
\includegraphics[width=7.5cm,height=4.6cm]{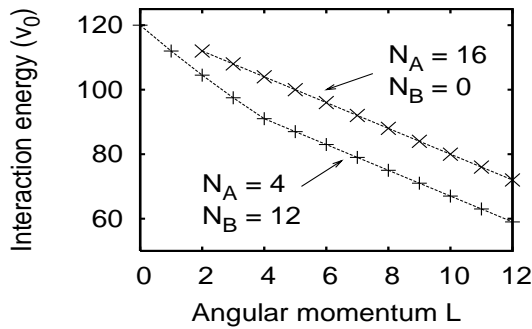}
\caption[]{The interaction energy that results from numerical
diagonalization of the Hamiltonian of a mixture of two Bose
gases, with $N_A = 4 $ and $N_B = 12$ (lower curve, marked by
``+"), as well as $N_A = 16$, and $N_B = 0$ (higher curve,
marked by ``x"), as function of the angular momentum $L$, for
$0 \le L \le 12$.}
\label{FIG1}
\end{figure}

\begin{figure}[t]
\includegraphics[width=6.5cm,height=8.1cm]{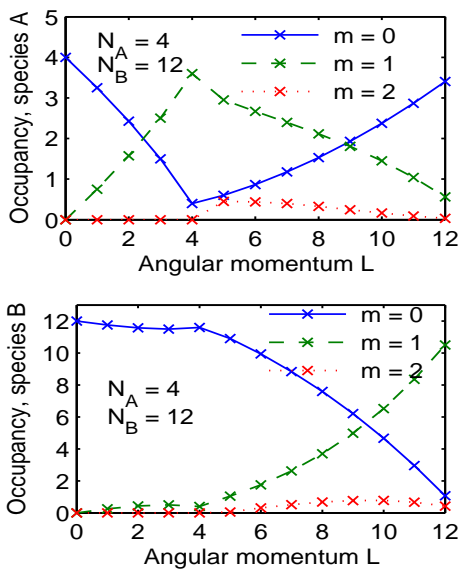}
\caption[]{The occupancy of the single-particle states with
$m=0, 1$ and 2, as function of the angular momentum $0 \le L
\le 12$, that results from numerical diagonalization of the
Hamiltonian of a mixture of two Bose gases, with $N_A = 4$ and
$N_B = 12$. The upper panel refers to species $A$, and the
lower one to species $B$.}
\label{FIG2}
\end{figure}

\begin{figure}[t]
\includegraphics[width=8.7cm,height=2.1cm,angle=-0]{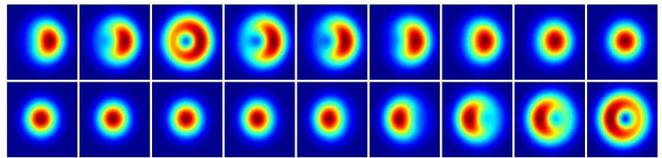}
\caption[]{The conditional probability distribution of a
mixture of two Bose gases, with $N_A = 4$ (higher row), and
$N_B = 12$ (lower row). Each graph extends between $-2.4 a_0$
and $2.4 a_0$. The reference point is located at $(x,y) =
(a_0,0)$ in the higher graph. The angular momentum $L$
increases from left to right, $L = 2, 3, 4 (=N_A), 5, 6, 8, 10,
11$, and $12 (= N_B)$.}
\label{FIG3}
\end{figure}

As $L$ becomes larger than $N_{\rm min}$, there is a phase
transition. For $N_{\rm min} \le L \le N_{\rm max}$, the
single-particle states that constitute the many-body state are
no longer only the ones with $m=0$ and $m=1$ (as in the case $0
\le L \le N_{\rm min}$), and in addition the interaction energy
varies linearly with $L$,
\begin{eqnarray}
  {\cal E}_0 (L)/v_0 = \frac 1 2 N (N-1)
  - \frac 1 4 N_{\rm min} N
  \nonumber \\
  - \frac 1 4 N L + \frac 1 4 N_{\rm min} (N_{\rm min}-1).
\end{eqnarray}
The lower curve in Fig.\,1 shows the interaction energy of a
system with $N_A=4$ and $N_B=12$, for $0 \le L \le 12$. For $0
\le L \le 4$ the energy is parabolic, and for $4 \le L \le 12$,
it is linear. These are exact results, within numerical
accuracy. The higher curve is the interaction energy of a
single-component system of $N=16$ atoms. It is known that in
this case, the interaction energy is exactly linear for $2 \le
L \le N = 16$ \cite{Bertsch}. This line is parallel to the line
which gives the interaction energy of the system with $N_A=4$
and $N_B=12$ for $N_A = 4 \le L \le N_B = 12$. Figure 2 shows
the occupancy of the single-particle states, that result from
the numerical diagonalization of the Hamiltonian.

The physical picture that emerges from these calculations is
intriguing: as $L$ increases, a vortex state enters the
component with the smaller population from infinity and ends up
at the center of the trap when $L = N_{\rm min}$. In addition,
another vortex state enters the component with the larger
population from the opposite side of the trap, reaching a
minimum distance from the center of the trap for $L \approx
N_A/2$ (this estimate is valid if $1 \ll N_A \ll N_B$), and
then returns to infinity when $L = N_{\rm min}$. This minimum
distance is $\approx 2 (N_B/N_A) a_0$. For $N_{\rm min} \le L
\le N_{\rm max}$, the vortex in the cloud with the smaller
population (that is located at the center of the trap when $L =
N_{\rm min}$) moves outwards, ending up at infinity when
$L=N_{\rm max}$, while a vortex in the other component moves
inwards again, ending up at the center of the trap when $L =
N_{\rm max}$. Figure 3 shows clearly these effects via the
conditional probability distributions, for $N_A=4$, and
$N_B=12$. We note that in the range $0 < L < N_A = 4$, the
(distant) vortex in the large component (lower row, species
$B$) is too far away from the center of the cloud to be
visible, because of the exponential drop of the density. The
plots in Fig.\,3 (and Fig.\,5) are not very sensitive to the
total number of atoms $N = N_A+N_B$, and resemble the behavior
of the system in the thermodynamic limit of large $N$.

In the case of equal populations, $N_A = N_B$, the parabolic
expression for the interaction energy, Eq.\,(\ref{yrast1}),
holds all the way between $0 \le L \le N_A = N_B$. Figure 4
shows the interaction energy and Fig.\,5 the occupancies of the
single-particle states, which vary linearly with $L$. The
corresponding physical picture is quite different in this case,
as shown in Fig.\, 6. The system is now symmetric with respect
to the two components, and a vortex state enters each of the
components (from opposite sides). These vortices reach a
minimum distance from the center of the trap equal to $a_0$,
when $L = N_A = N_B$.

\begin{figure}[t]
\includegraphics[width=7.5cm,height=4.6cm]{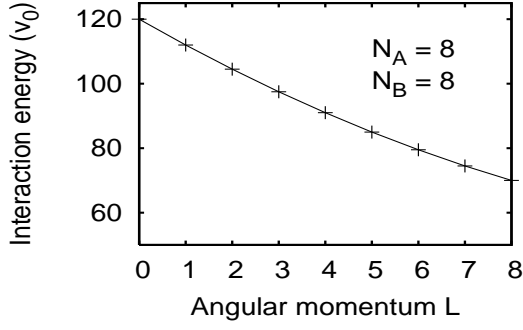}
\caption[]{The interaction energy that results from numerical
diagonalization of the Hamiltonian of a mixture of two Bose
 gases, with $N_A = N_B = 8$, as function of the angular
momentum $L$, for $0 \le L \le 8$.}
\label{FIG4}
\end{figure}

\begin{figure}[t]
\includegraphics[width=6.5cm,height=4.3cm]{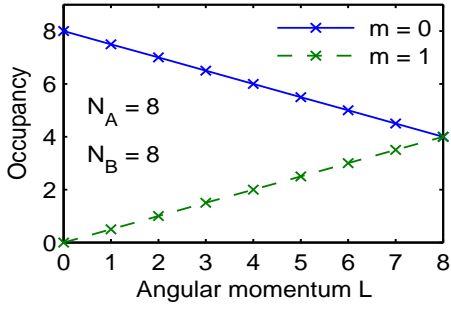}
\caption[]{The occupancy of the single-particle states with
$m=0$ and $m=1$, as function of the angular momentum $0 \le L
\le 8$, that results from numerical diagonalization of the
Hamiltonian of a mixture of two Bose gases, with $N_A = N_B =
8$.}
\label{FIG5}
\end{figure}

\begin{figure}[t]
\includegraphics[width=8.5cm,height=2.5cm]{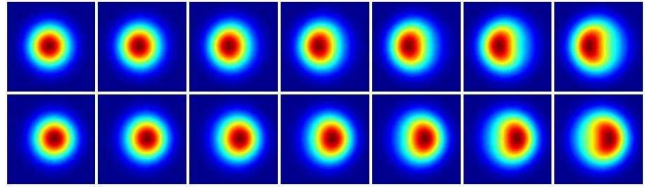}
\caption[]{The conditional probability distribution of a
mixture of two Bose gases, with $N_A = N_B = 8$. The two rows
refer to the two different species. Each graph extends between
$-2.4 a_0$ and $2.4 a_0$. The reference point is located at
$(x,y) = (a_0,0)$ in the lower graph. The angular momentum $L$
increases from left to right, $L = 2,3, \dots,8$.}
\label{FIG6}
\end{figure}

The simplicity of the system that we have studied allows one to
get some relatively simple analytical results, which we present
below. As we saw earlier, when $L=N_A$ or $L=N_B$, there is a
unit vortex state in species $A$ or $B$, while the other
species is in the lowest oscillator state with $m=0$. Since (at
least to leading order and next to leading order) only the
states with $m=0$ and $m=1$ are occupied, the Fock states are
of the general form (if, for example, $L=N_B$)
\begin{eqnarray}
 |n \rangle = |0^{N_A-n}, 1^n \rangle \bigotimes |0^n, 1^{N_B-n}
 \rangle.
\end{eqnarray}
Expressing the eigenstates of the interaction $V$ as $|\Phi
\rangle = \sum_n (-1)^n f_n | n \rangle$, the eigenvalue
equation takes the form
\begin{eqnarray}
  V_{n,n} f_n - V_{n,n-1} f_{n-1} - V_{n,n+1} f_{n+1}
  = {\cal E} f_n,
\end{eqnarray}
where $V_{n,m}$ are the matrix elements of the interaction
between the above states. Remarkably, if $N_A = N_B = N/2$,
then
\begin{eqnarray}
  V_{n,n} - V_{n,n-1} - V_{n,n+1} = 5 N (N-2) v_0/16,
\end{eqnarray}
which implies that in this case, the lowest eigenenergy is
${\cal E}_0 = 5 N (N-2) v_0/16$, in agreement with
Eq.\,(\ref{yrast1}). The corresponding eigenfunction is simply
$|\Phi_0 \rangle = \sum_n (-1)^n |n \rangle$.

In the case $N_A \neq N_B$, with, e.g., $L=N_B$, $n$ is of
order unity, and therefore the interaction may be written as
\begin{eqnarray}
  V/v_0 = \frac 1 2 N_{A} (N_{A}-1) + \frac 1 4 N_{B} (N_{B}-1)
  + \frac 1 2 N_{A} N_{B}
  + \nonumber \\
 +  \frac 1 2 (N_{A} + N_{B}) {\hat a}_0^{\dagger} {\hat a}_0
  + \frac 1 2 \sqrt {N_{A} N_{B}} ({\hat a}_0 {\hat b}_1^{\dagger} +
  {\hat a}_0^{\dagger} {\hat b}_1),
\end{eqnarray}
where ${\hat a}_m ({\hat a}_m^{\dagger})$ and ${\hat b}_m
({\hat b}_m^{\dagger})$ are annihilation (creation) operators
of the species $A$ and $B$ with angular momentum $m \hbar$. The
above expression for $V$ can be diagonalized with a Bogoliubov
transformation,
\begin{eqnarray}
   V/v_0 = \frac 1 4 |N_A - N_B| (2 {\hat \alpha}^{\dagger}
   {\hat \alpha} + 1)
  + \frac 1 4 N_A (2 N_A - 3) +
   \nonumber \\
   + \frac 1 4 N_B (N_B - 2) + \frac 1 2 N_A N_B,
\label{bog}
\end{eqnarray}
where ${\hat \alpha}^{\dagger} {\hat \alpha}$ is a number
operator. When $N_A = N_B = N/2$, the lowest eigenenergy is
$5N(N-2)/16$, in agreement with Eq.\,(\ref{yrast1}). When $N_A
\neq N_B$, the lowest eigenenergy is
\begin{eqnarray}
  {\cal E}_0 / v_0 = \frac 1 4 |N_A - N_B|
  + \frac 1 4 N_A (2 N_A - 3) +
  \nonumber \\
  + \frac 1 4 N_B (N_B - 2) + \frac 1 2 N_A N_B.
\label{enn}
\end{eqnarray}
The above expression agrees exactly with Eq.\,(\ref{yrast1})
when $N_A < N_B = L$, and to leading order in $N$ when $N_A >
N_B = L$.

In addition, according to Eq.\,(\ref{bog}), the excitation
energies are equally spaced, separated by $|N_A - N_B| v_0/2 +
{\cal O}(v_0)$. Therefore, one very important difference
between the case $N_A = N_B$ and $N_A \neq N_B$ is that in the
first case there are low-lying excited states, with an energy
separation of order $v_0$, while in the second [where in
general $N_A - N_B \sim {\cal O} (N)$], the low-lying excited
states are separated from the lowest state by an energy of
order $N v_0$.

Let us now turn to the mean-field description of this system,
for $0 \le L \le N_{\rm min}$. We consider the following order
parameters for the two species (restricting ourselves to the
states with $m=0$ and $m=1$ only),
\begin{equation}
  \Psi_A  = (c_{0} \Phi_0 + c_{1} \Phi_1) \varphi_0(z),
\nonumber \\
   \Psi_B = (d_{0} \Phi_0 + d_{1} \Phi_1) \varphi_0(z),
\end{equation}
where $c_0$, $c_1$, $d_0$, $d_1$ are variational parameters.
Given the order parameters, the many-body state is $\Phi_{\rm
MF} = \prod_{i=1}^{N_A} \Phi_A({\bf r}_i)  \prod_{j=1}^{N_B}
\Phi_B({\bf r}_j)$. The normalization for each species implies
that $|c_{0}|^2 + |c_{1}|^2 = 1$, and $|d_{0}|^2 + |d_{1}|^2 =
1$, while the condition for the angular momentum gives $N_A
|c_{1}|^2 + N_B |d_{1}|^2 = L$. The interaction energy is
\begin{eqnarray}
  {\cal E}_{\rm MF}  &=& \frac 1 2 N_A(N_A-1) U_0 \int |\Phi_A|^4
  \, d^3 r +
  \nonumber \\
   &+& \frac 1 2 N_B(N_B-1) U_0 \int |\Phi_B|^4 \, d^3 r +
\nonumber \\
   &+& N_A N_B U_0 \int |\Phi_A|^2  |\Phi_B|^2  \, d^3 r,
\end{eqnarray}
or, to leading order in $N$,
\begin{eqnarray}
  \frac {{\cal E}_{\rm MF}} {N^2 v_0} =
  \frac 1 2 \left( \frac {N_A} N \right)^2
  \left( |c_0|^4 + \frac 1 2|c_1|^4 + 2 |c_0|^2 |c_1|^2 \right) +
  \nonumber \\ +
  \frac 1 2 \left( \frac {N_B} N \right)^2
 \left( |d_0|^4 + \frac 1 2|d_1|^4 + 2 |d_0|^2 |d_1|^2 \right) +
\nonumber \\
+ \frac {N_A N_B} {N^2} \left( |c_0|^2 |d_0|^2 + \frac 1 2
|c_0|^2 |d_1|^2 + \frac 1 2 |c_1|^2 |d_0|^2 + \right. \nonumber \\ \left.
+ \frac 1 2 |c_1|^2 |d_1|^2 - |c_0| |c_1| |d_0| |d_1| \right),
\end{eqnarray}
where the phases of the variational coefficients have been
chosen so as to minimize ${\cal E}_{\rm MF}$.

When $N_A = N_B$, then $|c_0|^2 = |d_0|^2$, and also $|c_1|^2 =
|d_1|^2$, which implies that $|c_1|^2 = |d_1|^2 = L/N = l$.
Therefore, one finds that ${\cal E}_{\rm MF}/N^2 = (2 - 2 l +
{l^2}) v_0 / 4$, in agreement (to leading order in $N$) with
the result of exact diagonalization, Eq.\,(\ref{yrast1}). When
$N_A \neq N_B$, minimization of the energy with respect to one
of the four (free) variational parameters (the other three are
then fixed by the three constraints) gives a result that agrees
to leading order in $N$ with that of numerical diagonalization.

The fact that for $0 \le L \le N_{\rm min}$ the lowest
many-body state consists of only the $m=0$ and $m=1$
single-particle states is remarkable. To get some insight into
this result, we consider the two coupled Gross-Pitaevskii
equations, which describe the order parameters $\Psi_A$ and
$\Psi_B$. If $\mu_A$ and $\mu_B$ is the chemical potential of
each component, then
\begin{eqnarray}
 \left (- \frac {\hbar^2 \nabla^2} {2M} + V_t
 + U_0 |\Psi_B|^2 \right) \Psi_A
 + U_0 |\Psi_A|^2 \Psi_A &=& \mu_A \Psi_A,
 \nonumber \\
 \left(- \frac {\hbar^2 \nabla^2} {2M} + V_t
 + U_0 |\Psi_A|^2 \right) \Psi_B
+ U_0 |\Psi_B|^2 \Psi_B  &=& \mu_B \Psi_B.
\nonumber \\
\end{eqnarray}

For a large population imbalance, where, for example, $1 \ll
N_A \ll N_B$, for $0 \le L \le N_{\rm min} = N_A$, most of the
angular momentum is carried by the species $A$ (with the
smaller population). As mentioned earlier, in the range $0 \le
L \le N_{\rm min}$, although there is also a vortex state in
species $B$, this is far away from the center of the cloud. As
a result, the order parameter of species $B$ is essentially the
Gaussian state, with the corresponding density being
$|\Psi_B(\rho,z)|^2 \approx n_B(0,0)
e^{-\rho^2/a_0^2-z^2/a_z^2}$, where $n_B(0,0)$ is the density
of species $B$ at the center of the trap, i.e., at $\rho=0$ and
$z=0$.

This component acts as an external potential on species $A$.
Thus, the total ``effective" potential acting on species $A$ is
(expanding the function $e^{-\rho^2/a_0^2}$),
\begin{eqnarray}
  V_{\rm eff}(\rho,0)
   \approx \frac M 2 \omega^2 \rho^2
   + U_0 \, n_B(0,0)
   \left(1 - \frac {\rho^2} {a_0^2}
   + \frac {\rho^4} {2 a_0^4} \right),
\end{eqnarray}
for distances close to the center of the cloud. One may argue
that in this self-consistent analysis, the quadratic term in
the expansion changes the effective trap frequency, while the
quartic term acts as an anharmonic potential. We argue that
this anharmonic term is responsible for the fact that only the
states with $m=0$ and $m=1$ are occupied \cite{Emil}. Actually,
this is more or less how Dalibard {\it et al.} investigated the
problem of multiple quantization of vortex states
\cite{Dalibard}. In that case, it was an external laser beam
that created an external, repulsive Gaussian potential, as
opposed to the present problem, where this potential results
from the interaction between the different components.

We acknowledge financial support from the European Community
project ULTRA-1D (NMP4-CT-2003-505457), the Swedish Research
Council, and the Swedish Foundation for Strategic Research.

\end{document}